# Simulated-Annealing Optimization of Mean-Field Parameters in a Correlated-Basis Nuclear Model with Realistic Short-Range Correlations

Tianyang Ma[1] and Shalom Shlomo[1]

[1]*Cyclotron Institute, Texas A&M University, College Station, Texas, USA*

## Abstract

We optimize a six-parameter spherical Woods–Saxon mean field within a first-order correlated-basis-function description of finite nuclei containing central, spin–isospin, and tensor Short-Range Correlations (SRC). Monte Carlo importance sampling makes the multidimensional simulated-annealing search practical. The final objective contains 17 selected charge radii compared with a common theoretical scale of 0.030 fm. This light-to-heavy set is a representative compromise between mass–isospin coverage and the cost of repeatedly evaluating the correlated objective. Two independent fits are performed: S/S, in which the SRC theory uses its own fitted parameters, and 0/0, in which the no-SRC theory uses a separately fitted parameter set. A diagnostic S/0 calculation inserts the no-SRC-fitted parameters into the SRC theory. Its radius metric is 17.49 times the S/S value, demonstrating that the mean field must be refitted when explicit SRC are introduced. After separate optimization, radii, charge densities, and elastic form factors show no universal advantage for either consistent route. The corrected genuinely off-diagonal OBDM retains both uncorrelated kernels and the complete spectator-exchange spin contraction. In momentum space, the correlated S/S and diagnostic S/0 calculations generate the several-percent high-momentum strength observed in $^{12}$C and $^{40}$Ca, whereas the consistent no-SRC route 0/0 does not. The high-momentum tail is therefore the clearest signature of explicit SRC in the present calculation.

## 1. Introduction

Independent-particle mean-field models reproduce many global nuclear properties, but elastic electron scattering and nucleon-knockout measurements show that a single Slater determinant does not describe all features of the nuclear interior and spectral strength [1,2]. Correlations omitted from a pure mean-field treatment are commonly separated into long-range collective correlations and short-range correlations (SRC). SRC arise from the repulsive core and state-dependent tensor and spin–isospin components of the nucleon–nucleon interaction and generate structure at short interparticle distances and high momentum [12,34].

The correlated-basis-function (CBF) method constructs a correlated state by applying a state-dependent many-body correlation operator to an independent-particle reference state [3,33]. Direct evaluation becomes impractical as the number of particles and spin–isospin configurations increases. We therefore use a first-order cluster expansion following the diagrammatic approach of Iwamoto and Yamada [4]. The four retained diagram classes are hole/direct, hole/exchange, spectator/direct, and spectator/exchange [3–10].

A conceptual issue arises when phenomenological mean-field parameters fitted without explicit SRC are reused after an SRC operator is added. Part of the correlation effect can already be absorbed into the fitted one-body potential; adding an explicit correlation correction without refitting can therefore mix inconsistent model definitions [5]. Earlier Jastrow and CBF studies also show that correlated and uncorrelated calculations need not employ identical shell-model parameters [5,8].

The present work addresses this issue by optimizing the Woods–Saxon parameters independently in the correlated and uncorrelated theories. The final objective uses 17 selected charge radii. They were chosen as a middle-sized representative set spanning light through heavy nuclei and a useful range of neutron–proton asymmetries: using many more nuclei would substantially increase the cost of each correlated trial, whereas using too few would weaken the constraints on the mass and isospin dependence of the six parameters. Charge densities, elastic charge form factors, and momentum distributions are calculated afterward as validation observables. This design separates three questions: whether simulated annealing can optimize the mean field in the correlated model; whether explicit SRC require model-consistent refitting; and which observables retain a distinctive SRC signature after both theories have been fitted independently.

We use S/S for SRC theory with SRC-fitted parameters, 0/0 for no-SRC theory with no-SRC-fitted parameters, and S/0 for SRC theory evaluated with the no-SRC-fitted parameters. S/S and 0/0 are the consistent physical routes; S/0 is only a parameter-transfer diagnostic.

## 2. Correlated one-body density matrix and observables

### 2.1 Correlated state and first-order expansion

Before SRC are introduced, the reference state is the independent-particle Slater determinant

$$\Phi_0 = \frac{1}{\sqrt{A!}}\det[\phi_a(x_i)], \qquad a, i = 1, \ldots, A \tag{1}$$

The normalized correlated state and pair-product correlation operator are

$$\Psi = CF\Phi_0 \tag{2}$$

$$F = \prod_{i<j} f\,(i,j) \tag{3}$$

$$f(i,j) = 1 - g_c(r_{ij}) + f_{s\tau}(r_{ij})\sigma_i \cdot \sigma_j \tau_i \cdot \tau_j + f_{t\tau}(r_{ij}) S_{ij} \tau_i \cdot \tau_j \tag{4}$$

$$S_{ij} = 3(\sigma_i \cdot \hat{r}_{ij})(\sigma_j \cdot \hat{r}_{ij}) - \sigma_i \cdot \sigma_j \tag{5}$$

The functions $g_c$, $f_{s\tau}$, and $f_{t\tau}$ are given in Appendix A. The expansion is truncated at first order in the correlation line. Open circles in Fig. 1 are external coordinates, filled circles are integrated internal coordinates, solid lines are uncorrelated OBDM lines, and dashed lines are short-range correlation kernels [9,10].

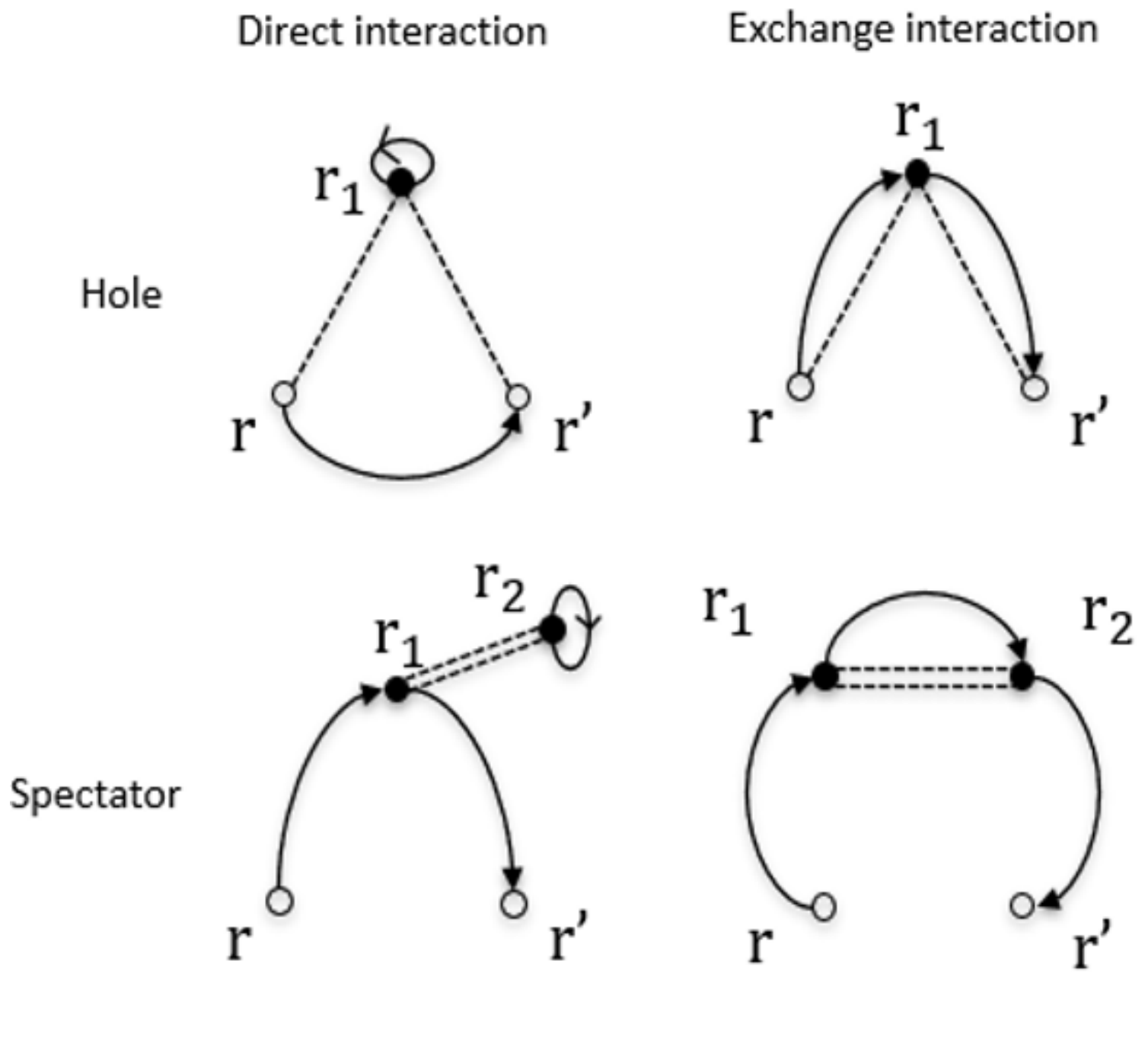

*Figure 1. First-order Mayer diagrams retained in the OBDM: hole/direct, hole/exchange, spectator/direct, and spectator/exchange.*

### 2.2 Uncorrelated spin-resolved OBDM

The radial single-particle functions are normalized as

$$\int_0^\infty 4\pi r^2 R_{nlj\tau}^2(r)dr = 1 \tag{6}$$

$$c = \hat{r}\cdot\hat{r}', C(r,r') = \hat{r}\times\hat{r}' \tag{7}$$

$$K_\tau(r,r') = \sum_{nlj\in\tau}\left(j+\frac{1}{2}\right)R_{nlj\tau}(r)R_{nlj\tau}(r')P_l(c) \tag{8}$$

$$Q_\tau(r,r') = \sum_{nlj\in\tau}(j-l)\,R_{nlj\tau}(r)R_{nlj\tau}(r')P_l'(c) \tag{9}$$

$$\rho_{0\tau\sigma\sigma'}(\mathbf{r},\mathbf{r}') = K_\tau\delta_{\sigma\sigma'} + iQ_\tau[\boldsymbol{\sigma}\cdot\mathbf{C}]_{\sigma\sigma'} \tag{10}$$

$$\rho_{0\tau\uparrow\uparrow} = K_\tau + iQ_\tau C_z, \rho_{0\tau\downarrow\downarrow} = K_\tau - iQ_\tau C_z \tag{11}$$

$$\rho_{0\tau\uparrow\downarrow} = iQ_\tau\left(C_x - iC_y\right), \rho_{0\tau\downarrow\uparrow} = iQ_\tau\left(C_x + iC_y\right) \tag{12}$$

$$\rho_{0\tau}^{sum}(r) = \sum_\sigma \rho_{0\tau\sigma\sigma}(r,r) = 2K_\tau(r,r) \tag{13}$$

For exchange terms define the two-line spin contraction

$$X_{\alpha\beta;\sigma\sigma'}(a,b,c) = \sum_{\sigma_1}\rho_{0\alpha\sigma\sigma_1}(a,b)\rho_{0\beta\sigma_1\sigma'}(b,c) \tag{14}$$

$$X_{\alpha\beta;\sigma\sigma'} = A_{\alpha\beta}\delta_{\sigma\sigma'} + i\left[\boldsymbol{\sigma}\cdot\mathbf{B}_{\alpha\beta}\right]_{\sigma\sigma'} \tag{15}$$

$$A_{\alpha\beta} = K_\alpha(a,b)K_\beta(b,c) - Q_\alpha(a,b)Q_\beta(b,c)\left[\left(\hat{a}\cdot\hat{b}\right)\left(\hat{b}\cdot\hat{c}\right) - \hat{a}\cdot\hat{c}\right] \tag{16}$$

$$B_{\alpha\beta} = K_\alpha Q_\beta\left(\hat{b}\times\hat{c}\right) + Q_\alpha K_\beta\left(\hat{a}\times\hat{b}\right) - Q_\alpha Q_\beta\left[\left(\hat{a}\times\hat{b}\right)\times\left(\hat{b}\times\hat{c}\right)\right] \tag{17}$$

The $Q_\tau$ and $B_{\alpha\beta}$ terms are retained at the OBDM level even when a complete local spin trace makes some of them vanish.

### 2.3 Separated-kernel structure

For a general N≠Z nucleus, the first-order OBDM is evaluated with nine separated scalar nonlocal kernels $H_1,\ldots,H_9$ and their diagonal one-vector limits $h_1,\ldots,h_9$. For a genuinely off-diagonal OBDM, the uncorrelated matrix $\rho_0\tau = K\tau I + iQ\tau\sigma\cdot C$ also couples to an antisymmetric spin-vector component generated by the tensor–tensor product in the direct sector. Define the relative vectors and the corresponding direct spin matrix by

$$x = r - y \tag{18}$$

$$x' = r' - y \tag{19}$$

$$c(x, x') = \hat{x} \cdot \hat{x}' \tag{20}$$

$$D(x, x') = \hat{x} \times \hat{x}' \tag{21}$$

$$\Xi(x, x') = 3c^2 - 1 \tag{22}$$

$$M_{\mathrm{i}}(x, x') = H_{\mathrm{i}}(x, x')I + iL_{\mathrm{i}}(x, x')\sigma \cdot D(x, x'), i = 1,2,3 \tag{23}$$

$$L_1 = L_2 = 9f_{\mathrm{t}}\tau(x)f_{\mathrm{t}}\tau(x')c(x, x') \tag{24}$$

$$L_3 = 36f_{\mathrm{t}}\tau(x)f_{\mathrm{t}}\tau(x')c(x, x') \tag{25}$$

$$½Tr_{\mathrm{s}}[\rho_0\tau(r, r')M_{\mathrm{i}}] = K\tau(r, r')H_{\mathrm{i}} - Q\tau(r, r')L_{\mathrm{i}}\Gamma \tag{26}$$

$$\Gamma = C(r, r') \cdot D(x, x') \tag{27}$$

The $H_i$ are the scalar kernels listed in Appendix B. The $L_i$ term is absent if the external OBDM is reduced to its spin-scalar kernel $K\tau$ alone. Because C(r,r)=0 and D_xy=0 when r′=r, the additional contribution vanishes identically in every local density, normalization, and radius calculation. The complete spin–isospin trace derivation, the orientation-sensitive proton–neutron contractions, and the combined $h_1$–$h_9$ plot are placed in Appendix B.

## 2.4 Exact first-order proton and neutron OBDMs

For compactness, define the two-line matrix product and the four-coordinate outer contraction

$$X_{\alpha\beta;ss'}(a, b, c) = \sum_{s_1} \rho_{0\alpha;ss_1}(a, b)\rho_{0\beta;s_1s'}(b, c) \tag{28}$$

$$\tilde{X}_{\alpha\gamma;ss'}(a, b; c, d) = \sum_{s_1} \rho_{0\alpha;ss_1}(a, b)\rho_{0\gamma;s_1s'}(c, d) \tag{29}$$

and the three-line contraction

$$Y_{\alpha\beta\gamma;ss'}(a, b, c, d) = \sum_{s_1s_2} \rho_{0\alpha;ss_1}(a, b)\rho_{0\beta;s_1s_2}(b, c) \times \rho_{0\gamma;s_2s'}(c, d) \tag{30}$$

The corrected first-order proton OBDM is the sum of the uncorrelated term and the four Mayer classes,

$$\rho_{p;ss'}(r,r') = \rho_{0p;ss'}(r,r') + A_{p;ss'} - B_{p;ss'} - C_{p;ss'} + D_{p;ss'} \tag{31}$$

where the hole/direct term is

$$\begin{gathered} A_{\mathrm{p};\,\mathrm{ss}}{'}(r,r') = 2\int d^3yK_{\mathrm{p}}(y,y)[\rho_{0\mathrm{p}}(r,r')M_1]_{\mathrm{ss}}{'} \\ +2\int d^3yK_{\mathrm{n}}(y,y)[\rho_{0\mathrm{p}}(r,r')M_2]_{\mathrm{ss}}{'} + 2\int d^3yK_{\mathrm{p}}(y,y)[\rho_{0\mathrm{n}}(r,r')M_3]_{\mathrm{ss}}{'} \end{gathered} \tag{32}$$

In Eq. (32), the products $\rho_{0\tau}M_i$ are spin-matrix products before the ss′ component is selected. Their spin trace gives Eq. (26). The scalar $H_i$ contribution reproduces the diagonal formalism, while the additional $-Q\tau L_i\Gamma$ term contributes only at finite separation. The hole/exchange term is

$$\begin{gathered} B_{p;ss'} = \int d^3yX_{pp;ss'}(r,y,r')H_4 + 8K_p(r,y)K_p(y,r')\delta_{ss'}H_5 \\ +2X_{pn;ss'}(r,y,r')H_6 + 16K_p(r,y)K_n(y,r')\delta_{ss'}H_7 \\ +2X_{np;ss'}(r,y,r')H_8 + 16K_n(r,y)K_p(y,r')\delta_{ss'}H_9 \end{gathered} \tag{33}$$

The spectator/direct term is

$$C_{p;ss'} = 2\int d^3yd^3zX_{pp;ss'}(r,y,r') \times \{K_p(z,z)h_1(r_{yz}) + K_n(z,z)[h_2(r_{yz}) + h_3(r_{yz})]\} \tag{34}$$

The different-isospin term in Eq. (34) is the spectator/direct trace used consistently in the density, radius, form-factor, and momentum routines. The spectator/exchange term is

$$\begin{gathered} D_{p;ss'} = \int d^3yd^3zY_{ppp;ss'}(r,y,z,r')h_4(r_{yz}) + 4\tilde{X}_{pp;ss'}(r,y;z,r')K_p(y,z)h_5(r_{yz}) \\ +2Y_{pnp;ss'}(r,y,z,r')[h_6(r_{yz}) + h_8(r_{yz})] \\ +8\tilde{X}_{pp;ss'}(r,y;z,r')K_n(y,z)[h_7(r_{yz}) + h_9(r_{yz})] \end{gathered} \tag{35}$$

The neutron OBDM is obtained by the explicit interchange $p \leftrightarrow n$, including reversal of the $pn$ and $np$ orientations:

$$\rho_{n;ss'} = \rho_{p;ss'}{}_{p\leftrightarrow n} \tag{36}$$

Equations (31)–(36), together with the matrix direct kernels in Eqs. (18)–(27), define the complete first-order proton OBDM used here. The neutron expression follows by p↔n with reversal of the pn and np orientations. The coefficients in the hole/exchange and spectator/exchange terms arise from different spin contractions and diagrammatic multiplicities. The three-line spectator/exchange contraction retains the KKK term, all three KQQ terms, and the cubic QQQ orientation term. These are physical OBDM contributions and are not altered by the Monte Carlo proposal distribution.

## 2.5 Charge radius, charge density, and form factor

$$\rho_\tau(r) = \sum_\sigma \rho_{\tau\sigma\sigma}(r,r), \int d^3r\rho_p = Z, \int d^3r\rho_n = N \tag{37}$$

$$\langle r^2\rangle_p = \frac{1}{Z}\int d^3rr^2\rho_p(r) \tag{38}$$

$$R_{ch}^2 = \left(1-\frac{2}{A}\right)\langle r^2\rangle_p + \frac{1}{A}\langle r^2\rangle_m + 0.707 - 0.116\frac{N}{Z} + 0.12533 s_p - 0.08585 s_n \tag{39}$$

The first two terms implement the fixed approximate center-of-mass correction used in both the fit and the final density/form-factor codes [11]. The constants 0.707 and −0.116 fm² are the proton and neutron finite-size contributions used in the code, while $s_p$ and $s_n$ are the corresponding spin–orbit factors.

$$F_\tau(q) = 4\pi \int_0^\infty r^2 j_0(qr)\rho_\tau(r)dr \tag{40}$$

$$G_{Ep}(q) = exp(-0.707\, q^2/6) \quad G_{En}(q) = \frac{0.116q^2}{6} G_{Ep}(q) \tag{41}$$

$$F_{ch}(q) = \frac{1}{Z}\left[G_{Ep}(q)F_p(q) + G_{En}(q)F_n(q)\right] - \frac{q^2}{6} G_{Ep}(q)\left\{0.12533 F_{\eta p}(q) - 0.08585 F_{\eta n}(q)\right\} \tag{42}$$

$$F_{\rm ch}(0) = 1, \qquad R_{\rm ch}^2 = -6\frac{dF_{\rm ch}}{dq^2}\bigg|_{q=0} \tag{43}$$

The same neutron spin–orbit normalization is used in the radius and density codes, so the neutron term remains $-0.08585 s_n$ rather than acquiring an additional N/Z factor.

### 2.6 Momentum distribution

The momentum distribution is obtained from the genuinely off-diagonal OBDM. Define

$$u = t + \frac{s}{2}, v = t - \frac{s}{2} \tag{44}$$

$$F_\tau(s) = \sum_s \int d^3t \rho_{\tau;ss}(u,v) \tag{45}$$

Using Eq. (31), the finite off-diagonal trace is decomposed into the same four first-order classes,

$$F_\tau = F_{0,\tau} + F_{A,\tau} - F_{B,\tau} - F_{C,\tau} + F_{D,\tau} \tag{46}$$

$$F_{0,\tau}(s) = \sum_s \int d^3t \rho_{0\tau;ss}(u,v) \tag{47}$$

$$F_{A,\tau}(s) = \sum_s \int d^3t A_{\tau;ss}(u,v) \tag{48}$$

$$F_{B,\tau}(s) = \sum_s \int d^3t B_{\tau;ss}(u,v) \tag{49}$$

$$F_{C,\tau}(s) = \sum_{s} \int d^3t C_{\tau;ss}(u, v) \quad (50)$$

$$F_{D,\tau}(s) = \sum_{s} \int d^3t D_{\tau;ss}(u, v) \quad (51)$$

The explicit integrands in Eqs. (48)–(51) are obtained by substituting Eqs. (32)–(35). The finite-separation calculation retains the scalar $H_1$–$H_9$ kernels, the direct spin-vector kernels $L_1$–$L_3$, both pn/np orientations, and the complete three-line Q-dependent contraction before the final spin trace. The new direct term is proportional to Γ and therefore vanishes exactly at zero separation. The Fourier transform is

$$n_\tau(k) = \frac{1}{(2\pi)^3} \int d^3s e^{ik\cdot s} F_\tau(s) \quad (52)$$

$$n_\tau(k) = \frac{1}{2\pi^2} \int_0^\infty s^2\, ds j_0(ks) F_\tau(s) \quad (53)$$

$$4\pi \int_0^\infty k^2\, n_\tau(k) dk = N_\tau \quad (54)$$

For numerical evaluation, the half-separation variable is used:

$$G_\tau(\delta) = \sum_{s} \int d^3t \rho_{\tau;ss}(t+\delta, t-\delta) = F_\tau(2\delta) \quad (55)$$

$$n_\tau(k) = \frac{4}{\pi^2} \int_0^\infty \delta^2\, d\delta j_0(2k\delta) G_\tau(\delta) \quad (56)$$

The production SRC output uses the stable difference construction

$$n_\tau^{\mathrm{SRC}}(k) = n_{\tau,0}^{\mathrm{direct}}(k) + \mathcal{F}\left[G_\tau^{\mathrm{SRC}}(\delta) - G_{\tau,0}(\delta)\right] \quad (57)$$

The direct no-SRC transform is evaluated deterministically on the same momentum grid. For the correlated calculation, the Monte Carlo samples are reused across all half-separation points in a common random block, reducing artificial point-to-point noise before the Fourier–Bessel transform. The local value $G_\tau(0)$ is checked against the particle number before any final numerical normalization. The corrected nonlocal SRC codes retain the full $Q_\tau$-dependent direct and spectator-exchange structures described above. The final momentum comparisons will be regenerated after these codes complete their production runs.

## 3. Independent optimization of Woods–Saxon parameters

### 3.1 Woods–Saxon Hamiltonian

$$H_\tau = -\frac{\hbar^2}{2\mu_\tau}\nabla^2 + V_\tau(r) + \delta_{\tau p}V_C(r) + \frac{1}{r}\frac{dV_{SO}(r)}{dr}l \cdot s \tag{58}$$

$$V_\tau(r) = -V_\tau f(r, R, a), V_{SO}(r) = \lambda V_0 f(r, R_{SO}, a) \tag{59}$$

$$f(r, R, a) = \frac{1}{1 + \exp[(r - R)/a]}, \quad R = r_0 A^{1/3}, \quad R_{SO} = r_{SO} A^{1/3} \tag{60}$$

$$V_\tau = V_0\left[1 - \frac{4\kappa}{A}\langle \mathbf{t} \cdot \mathbf{T}' \rangle\right] \tag{61}$$

The Coulomb term uses a uniformly charged sphere with core charge $Z - 1$. The spin–orbit term is not multiplied by the central isospin factor. The functional form, starting point, and search ranges follow established Woods–Saxon parameterizations used in shell-model and microscopic–macroscopic calculations [14–20]. The six fitted parameters are

$$v = (V_0, r_0, r_{SO}, a, \kappa, \lambda) \tag{62}$$

*Table 1. Starting point and numerical search limits used in both fits.*

| Parameter | Unit | Starting value | Allowed interval | Maximum trial displacement |
|---|---|---|---|---|
| $V_0$ | MeV | 52.06 | 40.0–60.0 | 4 |
| $r_0$ | fm | 1.26 | 1.10–1.50 | 0.05 |
| $r_{SO}$ | fm | 1.16 | 0.90–1.50 | 0.1 |
| a | fm | 0.662 | 0.30–1.20 | 0.1 |
| κ | — | 0.639 | 0.40–0.90 | 0.1 |
| λ | — | 24.1 | 15.0–40.0 | 2 |

### 3.2 Radius-only objective

$$\chi_R^2(v) = \sum_{i=1}^{17}\left[\frac{R_{ch,i}^{th}(v) - R_{ch,i}^{exp}}{\sigma_R}\right]^2, \quad \sigma_R = 0.030 \text{ fm} \tag{63}$$

$$\begin{aligned}\chi_\nu^2(v) &= \frac{1}{N_d - N_p}\sum_{i=1}^{N_d}\left[\frac{R_{ch,i}^{th}(v) - R_{ch,i}^{exp}}{\sigma_R}\right]^2 \\ &= \frac{1}{11}\sum_{i=1}^{17}\left[\frac{R_{ch,i}^{th}(v) - R_{ch,i}^{exp}}{\sigma_R}\right]^2\end{aligned} \tag{64}$$

The same objective definition is used in the two independent searches, but the parameter vectors are not shared. The 17-nucleus set is a deliberate computational and statistical compromise rather than a claim that only 17 charge radii are available: it provides representative light-to-heavy and isospin coverage while keeping each Monte Carlo objective evaluation practical.

## 3.3 Simulated annealing and screening

$$\Delta\chi^2 = \chi_R^2(v') - \chi_R^2(v) \tag{65}$$

$$P_{\text{acc}} = \min[1, \exp(-\Delta\chi^2/T)] \tag{66}$$

$$T_i = \frac{2}{i}, i = 1, \dots, 50 \tag{67}$$

The production schedule contains 50 temperatures, T=2/i for i=1,…,50. Rounds 1–15, 16–30, and 31–50 use 10,000, 20,000, and 100,000 Monte Carlo samples per radius evaluation, respectively. The code stores both the current and historical-best parameter vectors and can resume from the last completed temperature. A six-nucleus prescreen ($^{12}$C, $^{16}$O, $^{40}$Ca, $^{48}$Ca, $^{60}$Ni, and $^{90}$Zr) rejects clearly inferior trials early, but every accepted candidate and every reported objective is evaluated with all 17 nuclei. Because the SRC objective is a Monte Carlo estimator, a low stored search value is treated as a candidate, not as the final fit quality. Final parameters and radius metrics are taken from the separately checked production outputs; the temperature history documents the search rather than asymptotic convergence.

# 4. Experimental inputs and validation references

### 4.1 Charge radii used in the fit

The 17 fitted charge radii are listed in Table 2. They were selected as a representative middle-sized sample spanning light, medium, and heavy systems and both near-symmetric and neutron-rich nuclei. Adding many more nuclei would increase the repeated correlated OBDM cost, while a much smaller set would weaken the constraints on the six Woods–Saxon parameters. Most values are taken from the compilation of Angeli and Marinova [25]. More recent measurements are used for $^{36}$Ca [26], $^{56}$Ni [27], and $^{68}$Ni [28]. The quoted experimental uncertainties are reported for reference, whereas the optimization uses the common theoretical fitting uncertainty σR=0.030 fm for every nucleus.

*Table 2. Charge-radius inputs.*

| Nucleus | $R_{\text{ch}}$ (fm) | Quoted uncertainty (fm) | Source |
|---|---|---|---|
| $^{12}$C | 2.4702 | 0.0022 | [25] |
| $^{14}$C | 2.5025 | 0.0087 | [25] |
| $^{16}$O | 2.7013 | 0.0052 | [25] |

| Nucleus | $R_{ch}$ (fm) | Quoted uncertainty (fm) | Source |
|---|---|---|---|
| $^{28}Si$ | 3.1224 | 0.0024 | [25] |
| $^{30}Si$ | 3.1336 | 0.004 | [25] |
| $^{32}S$ | 3.2608 | 0.0018 | [25] |
| $^{36}S$ | 3.2985 | 0.0024 | [25] |
| $^{36}Ca$ | 3.45 | 0.01 | [26] |
| $^{40}Ca$ | 3.478 | 0.0019 | [25] |
| $^{48}Ca$ | 3.478 | 0.002 | [25] |
| $^{56}Ni$ | 3.723 | 0.005 | [27] |
| $^{60}Ni$ | 3.806 | 0.0016 | [25] |
| $^{68}Ni$ | 3.886 | 0.005 | [28] |
| $^{88}Sr$ | 4.224 | 0.0018 | [25] |
| $^{90}Zr$ | 4.27 | 0.001 | [25] |
| $^{140}Ce$ | 4.8771 | 0.0018 | [25] |
| $^{208}Pb$ | 5.5012 | 0.0013 | [25] |

### 4.2 Charge-density and form-factor shape references

Historical Fourier–Bessel (FB) charge-density reconstructions from elastic electron scattering are taken from the de Vries–de Jager–de Vries compilation and associated UVA coefficient files [31]. These reconstructions are not additional fit data. Their raw rms radii differ slightly from the recommended charge radii used in Table 2, so the FB reference is radius matched before the shape comparison. With $\lambda = R_{FB}/R_{fit}$,

$$\rho_{FB}^{matched}(r) = \lambda^3 \rho_{FB}(\lambda r) \tag{68}$$

$$F_{FB}^{matched}(q) = F_{FB}(q/\lambda) \tag{69}$$

This transformation preserves normalization and gives the trusted fit radius. It is used only as a shape diagnostic. The matched FB curve is shown only within the momentum-transfer support of the original reconstruction; beyond that range the figures show theory only.

### 4.3 Momentum-distribution references

The absolute $^{12}C$ and $^{40}Ca$ per-nucleon momentum distributions are compared with the inclusive-scattering extraction of Liang, Bai, and Ren [32]. These distributions are validation observables and do not enter the fit. The experimental point ranges are used for the displayed comparisons. Integrated high-momentum probabilities are interpreted as scale comparisons rather than a common-window $\chi^2$ because the measured upper limits differ between the two nuclei and from the common theoretical integration limit.

## 5. Results

## 5.1 Simulated-annealing history

The correlated and uncorrelated Woods–Saxon searches were annealed independently with the same radius-only objective and the common scale $\sigma_R$=0.030 fm. Figure 2 reproduces the approved constructed $\chi^2_R$ histories for the two routes. The curves are search diagnostics: they show the reduction and stabilization of the objective as 1/T increases, but they are not used as a claim of asymptotic Monte Carlo convergence.

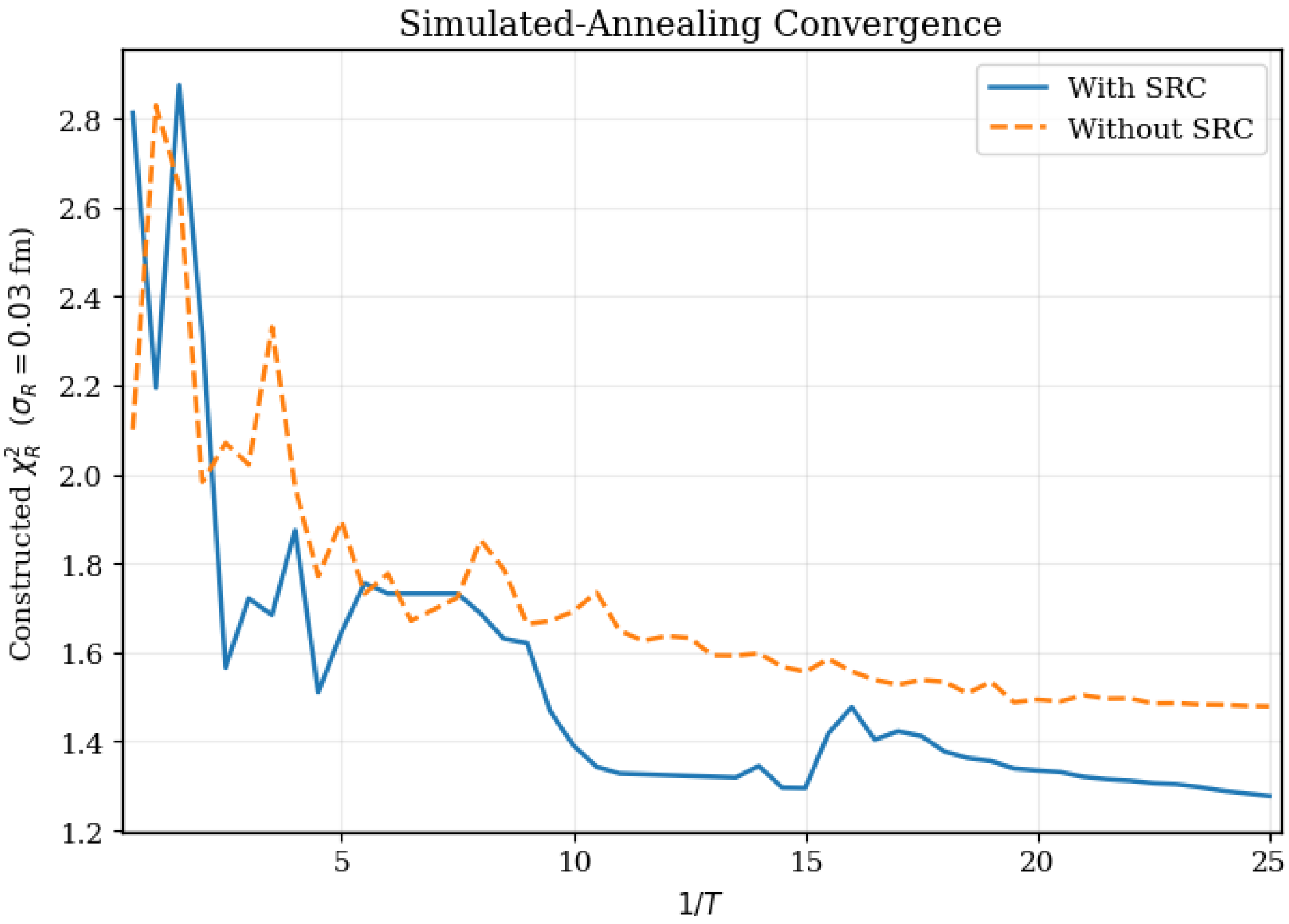

**Figure 2. Simulated-annealing convergence diagnostic. Constructed radius objective $\chi^2_R$ versus inverse temperature for the independently annealed SRC and no-SRC routes, using $\sigma_R$=0.030 fm.**

The separate $\Delta\chi^2$ diagnostic is shown in Figure 3. Both archived sequences, labelled ν and $\nu_1$ in the original plotting record, fall rapidly toward zero as the temperature decreases, indicating that accepted changes become progressively smaller late in the schedules.

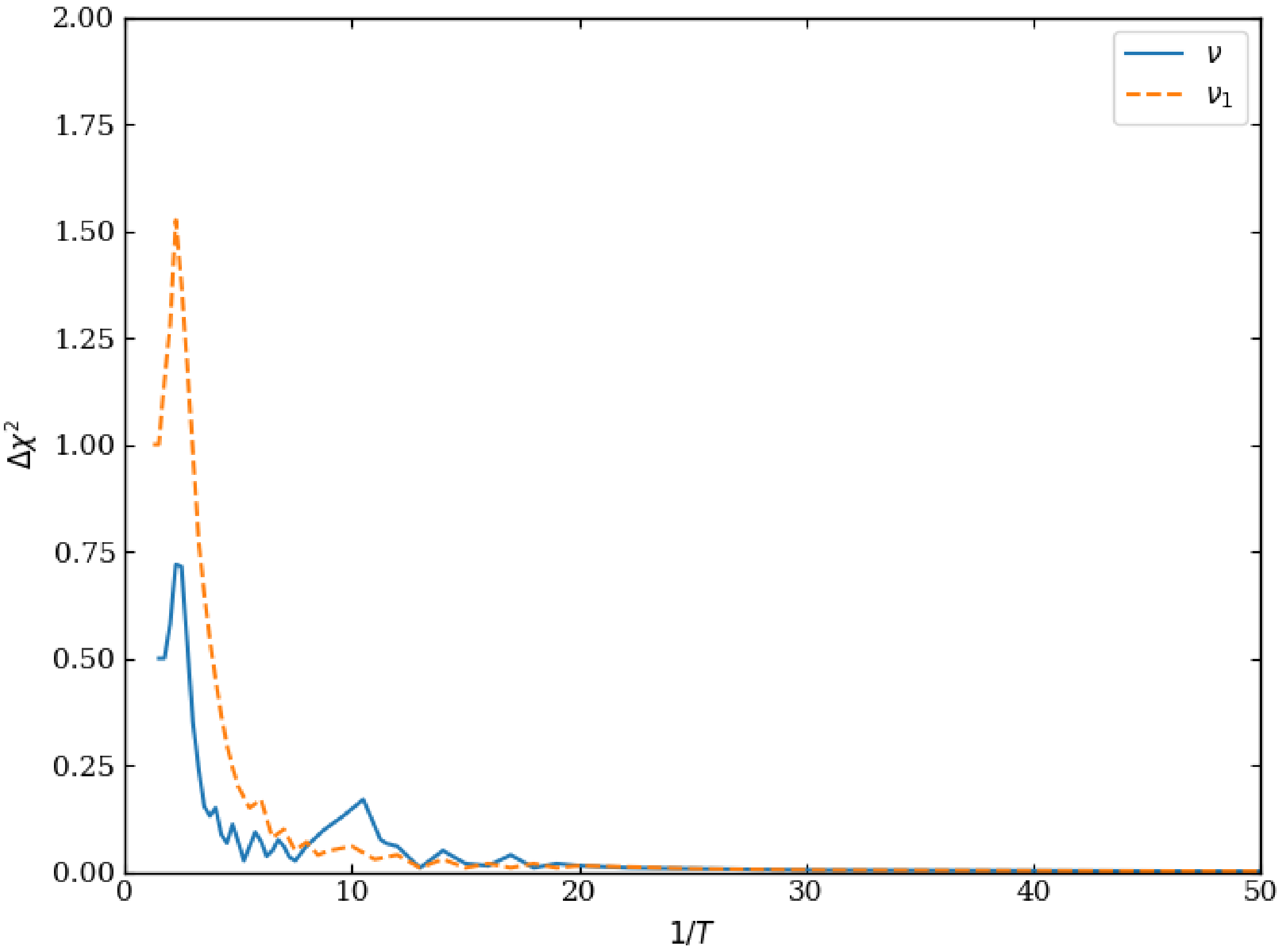

Figure 3. Change in the constructed radius objective, $\Delta\chi^2$, versus inverse temperature. The solid ν and dashed $\nu_1$ curves retain the labels of the approved archived diagnostic.

## 5.2 Fitted Woods–Saxon parameters

Table 3. Independently fitted Woods–Saxon parameters and the corrected final reduced radius metrics.

| Parameter | S/S: SRC-fitted | 0/0: no-SRC-fitted |
|---|---|---|
| $V_0$ | 56.3632 | 53.0146 |
| $r_0$ | 1.3379 | 1.2819 |
| $r_{SO}$ | 0.9466 | 0.9625 |
| $a$ | 0.9928 | 0.7404 |
| $\kappa$ | 0.4005 | 0.4009 |
| $\lambda$ | 39.8268 | 35.8881 |
| $\chi_{red}^2$ | 1.2059 | 1.4786 |

The sizable changes in $r_0$, diffuseness a, and spin–orbit strength λ show that the effective one-body field reorganizes when explicit SRC are introduced. The parameter values themselves are not observables; the final $\chi^2_{red}$ row is obtained from the corrected 17-radius check with ν=17−6=11.

## 5.3 Charge radii

*Table 4. Experimental and calculated charge radii.*

| Nucleus | Experiment (fm) | S/S (fm) | S/S residual (fm) | 0/0 (fm) | 0/0 residual (fm) |
|---|---|---|---|---|---|
| $^{12}$C | 2.4702 | 2.51367 | 0.043471 | 2.50204 | 0.031841 |
| $^{14}$C | 2.5025 | 2.51947 | 0.016974 | 2.51032 | 0.007823 |
| $^{16}$O | 2.7013 | 2.69223 | -0.009072 | 2.68417 | -0.01713 |
| $^{28}$Si | 3.1224 | 3.12305 | 0.00065 | 3.11336 | -0.009043 |
| $^{30}$Si | 3.1336 | 3.13197 | -0.001632 | 3.11673 | -0.016874 |
| $^{32}$S | 3.2608 | 3.24335 | -0.017451 | 3.24313 | -0.01767 |
| $^{36}$S | 3.2985 | 3.25231 | -0.046192 | 3.25318 | -0.04532 |
| $^{36}$Ca | 3.45 | 3.46367 | 0.013666 | 3.49483 | 0.044832 |
| $^{40}$Ca | 3.478 | 3.44679 | -0.031209 | 3.449 | -0.029003 |
| $^{48}$Ca | 3.478 | 3.44735 | -0.030646 | 3.44927 | -0.028729 |
| $^{56}$Ni | 3.723 | 3.77336 | 0.050356 | 3.7933 | 0.070296 |
| $^{60}$Ni | 3.806 | 3.80138 | -0.004621 | 3.80658 | 0.000576 |
| $^{68}$Ni | 3.886 | 3.84864 | -0.037359 | 3.85852 | -0.027478 |
| $^{88}$Sr | 4.224 | 4.20963 | -0.014369 | 4.215 | -0.009004 |
| $^{90}$Zr | 4.27 | 4.25906 | -0.010935 | 4.26399 | -0.006013 |
| $^{140}$Ce | 4.8771 | 4.89449 | 0.017385 | 4.89937 | 0.022267 |
| $^{208}$Pb | 5.5012 | 5.52477 | 0.023566 | 5.52467 | 0.02347 |

The rms residual is 0.02650 fm for S/S and 0.02934 fm for 0/0. Both independently fitted models reproduce the overall radius systematics at the common 0.030 fm comparison scale.

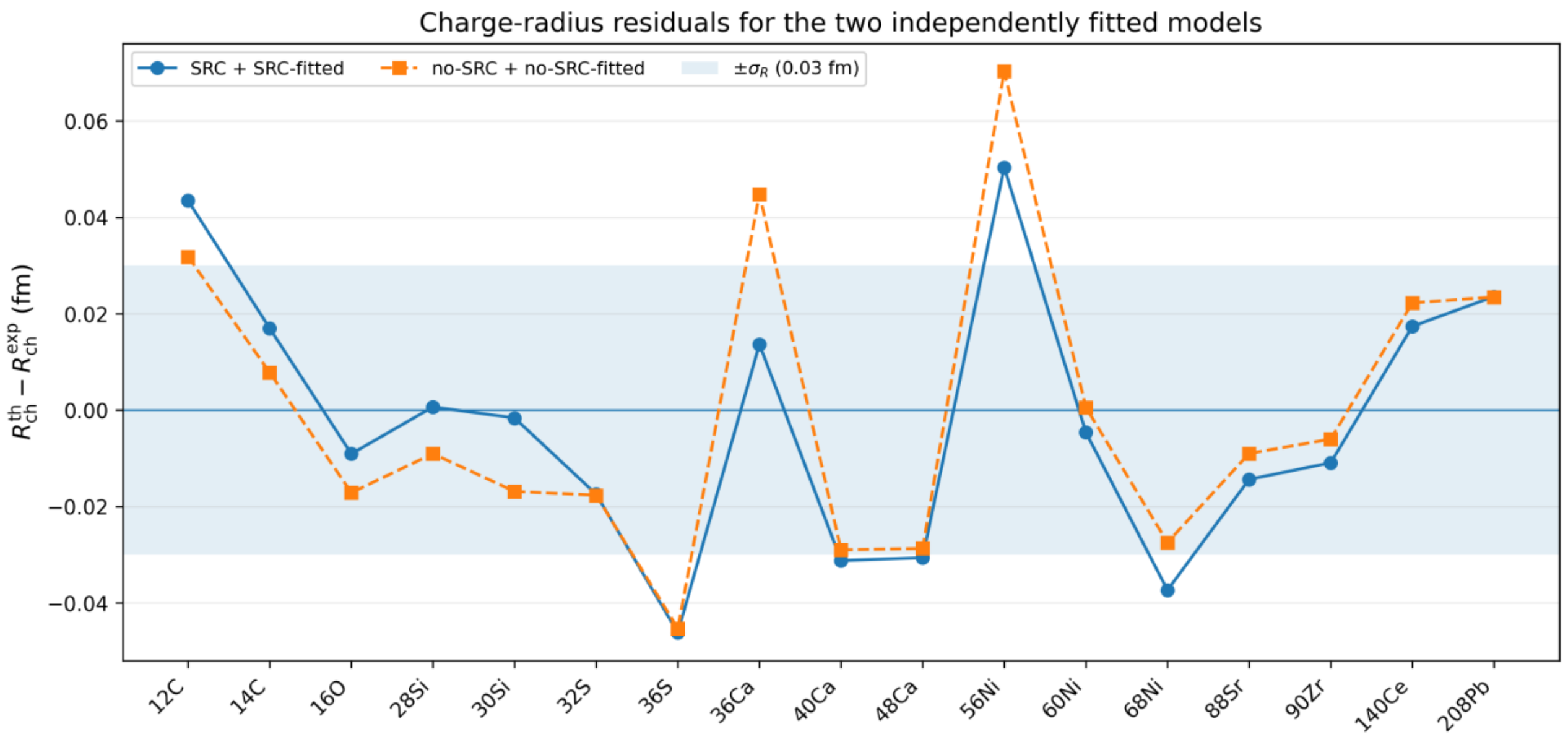

Figure 4. Charge-radius residuals after independent SRC and no-SRC optimization. The shaded band shows ±0.03 fm.

## 5.4 Necessity of model-consistent refitting

*Table 5. Radius-only fit-quality comparison. S/0 is a diagnostic, not a physical fitted prediction.*

| Calculation | Raw metric | Reduced metric (÷11) | Role |
|---|---|---|---|
| S/S: SRC theory + SRC-fitted parameters | 13.2644 | 1.20585 | consistent SRC result |
| 0/0: no-SRC theory + no-SRC-fitted parameters | 16.2649 | 1.47863 | consistent no-SRC baseline |
| S/0: SRC theory + no-SRC-fitted parameters | 232.021 | 21.0928 | diagnostic only |

The S/0 raw metric is 17.492 times the S/S metric. Because both values are evaluated in the same correlated theory with the same radii and weighting, the difference is caused by transferring a parameter set fitted in the no-SRC model into the SRC model. The numerical results therefore demonstrate the necessity of model-consistent refitting within the present SRC framework. This is not a claim of a universal mathematical proof, nor does it require S/S to outperform 0/0 for every observable.

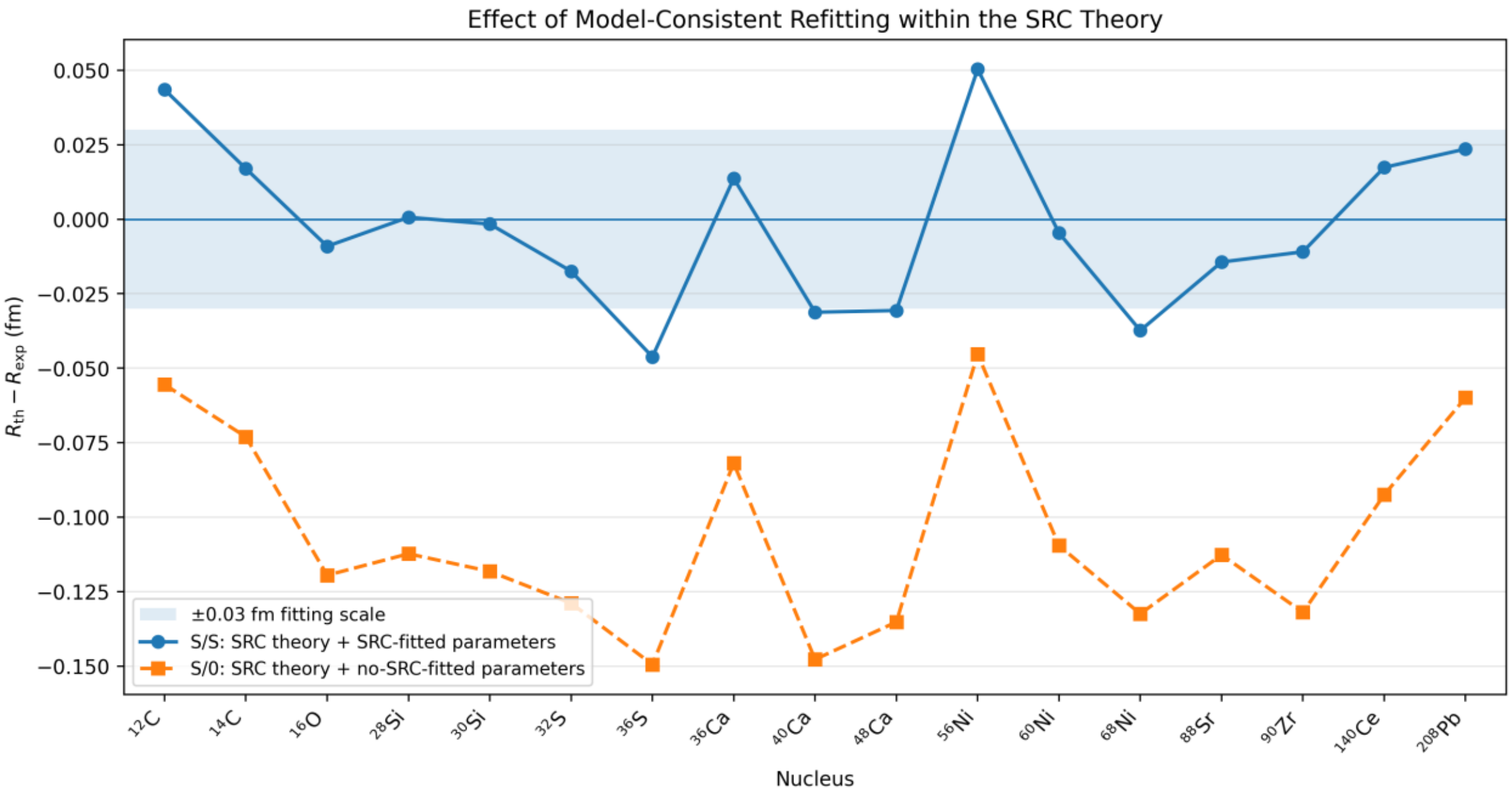

Figure 5. Necessity of model-consistent refitting within the corrected SRC theory. S/S uses the SRC-fitted parameters; S/0 inserts the no-SRC-fitted parameters into the same SRC theory. The shaded band is ±0.03 fm. S/0 is a diagnostic, not a third fitted model.

## 5.5 Charge densities

Before comparison with electron-scattering reconstructions, the independently implemented density and form-factor routines were checked against the radius calculations. The proton and neutron densities reproduce Z and N, the charge form factor satisfies $F_{ch}(0)=1$, and the rms radius obtained by integrating the final charge density agrees with the direct radius calculation. For the deterministic no-SRC calculation the differences are $2.0\times10^{-5}$–$4.3\times10^{-5}$ fm for the six plotted nuclei. For the SRC calculation, three independent production seeds give density-integrated and direct radii consistent within 1.6 combined Monte Carlo standard errors. The plotted SRC curves are the averages of three independent 300,000-sample unsmoothed runs.

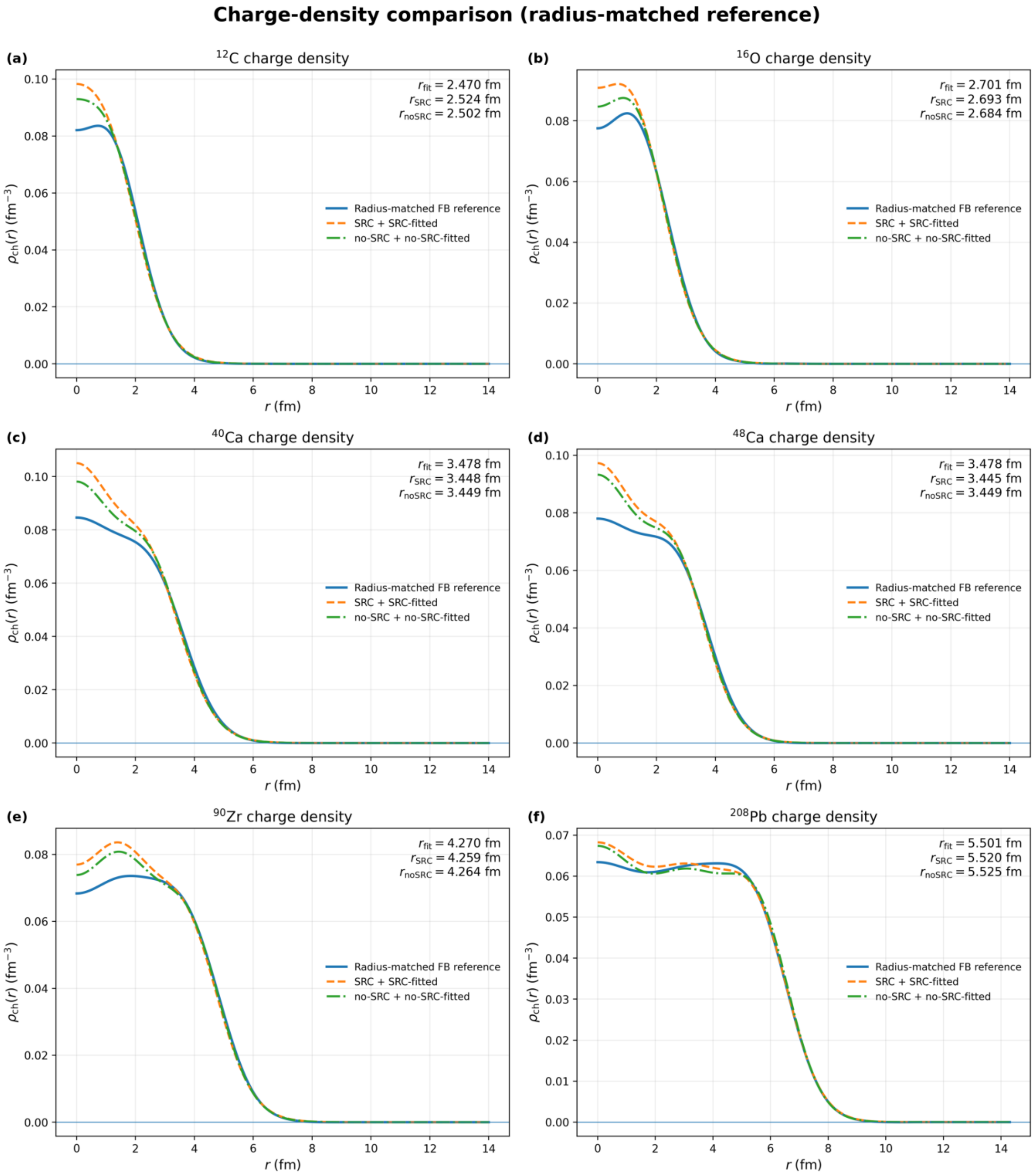

Figure 6. Radius-matched charge-density comparison for six nuclei. The historical Fourier–Bessel reference is from the de Vries compilation [31]; only the matched reference, S/S, and 0/0 are shown.

After the historical FB reference is matched to the same rms charge radius used in the fit, both S/S and 0/0 give generally reasonable interior, surface, and tail shapes. The comparison does not reveal a systematic coordinate-space advantage for S/S across all six nuclei. This result is consistent with the purpose of the separate fits: part of the SRC-induced redistribution can be absorbed into a refitted effective mean field.

## 5.6 Elastic charge form factors

Figure 7 presents the single form-factor set used for the coordinate-space comparison. All panels plot $|F_{ch}(q)|^2$ on a logarithmic ordinate. The displayed momentum-transfer range is 0–6 $fm^{-1}$ for $^{12}C$, 0–8 $fm^{-1}$ for $^{16}O$, $^{40}Ca$, $^{48}Ca$, and $^{90}Zr$, and 0–10 $fm^{-1}$ for $^{208}Pb$. The lower ordinate limit is adjusted separately for each nucleus so that the calculated diffraction structure remains visible over the full requested range.

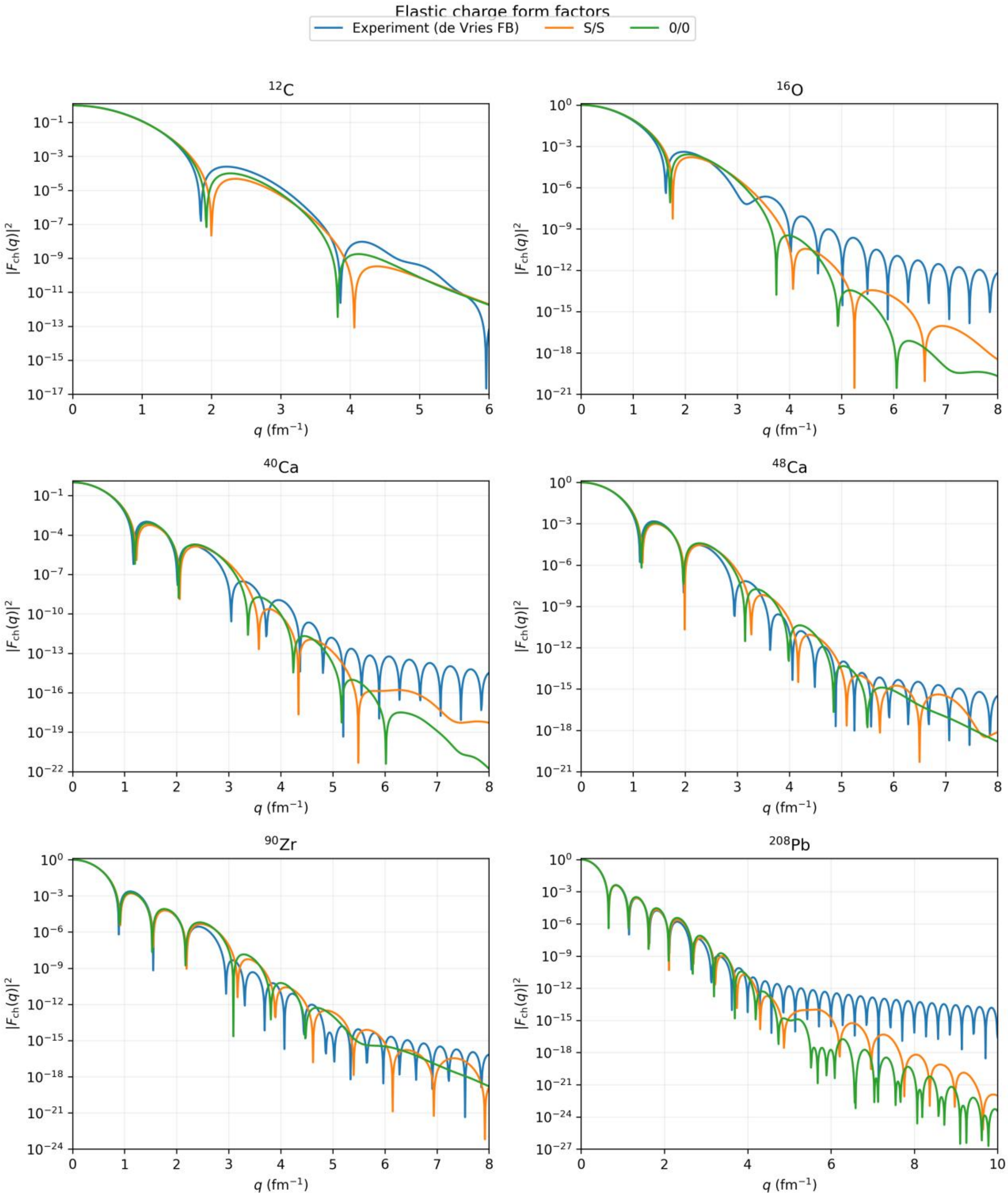

**Figure 7. Final elastic charge-form-factor comparison. The radius-matched de Vries Fourier–Bessel reference [31], S/S, and 0/0 are shown as $|F_{ch}(q)|^2$ on logarithmic scales. The q ranges are 6 fm$^{-1}$ for $^{12}$C, 8 fm$^{-1}$ for $^{16}$O, $^{40}$Ca, $^{48}$Ca, and $^{90}$Zr, and 10 fm$^{-1}$ for $^{208}$Pb.**

Both independently fitted routes reproduce the principal low-q diffraction pattern, but the higher minima and oscillations remain nucleus dependent. Across the six nuclei, the form-factor comparison is mixed rather than uniformly improved by SRC; the clearest S/S advantage occurs for the $^{208}$Pb form-factor shape.

## 5.7 Momentum distributions and the explicit SRC signal

The momentum distribution probes the genuinely off-diagonal OBDM and therefore tests physics that cannot be absorbed entirely into a smooth refitted mean field. Figure 8 compares the corrected absolute per-nucleon distributions for $^{12}$C and $^{40}$Ca with the digitized inclusive-scattering extraction of Liang, Bai, and Ren [32]. S/S and 0/0 are the two consistently fitted routes. S/0 is retained only as a parameter-transfer diagnostic. Both correlated calculations produce a broad high-k tail, whereas the deterministic 0/0 result falls rapidly once the mean-field region is left.

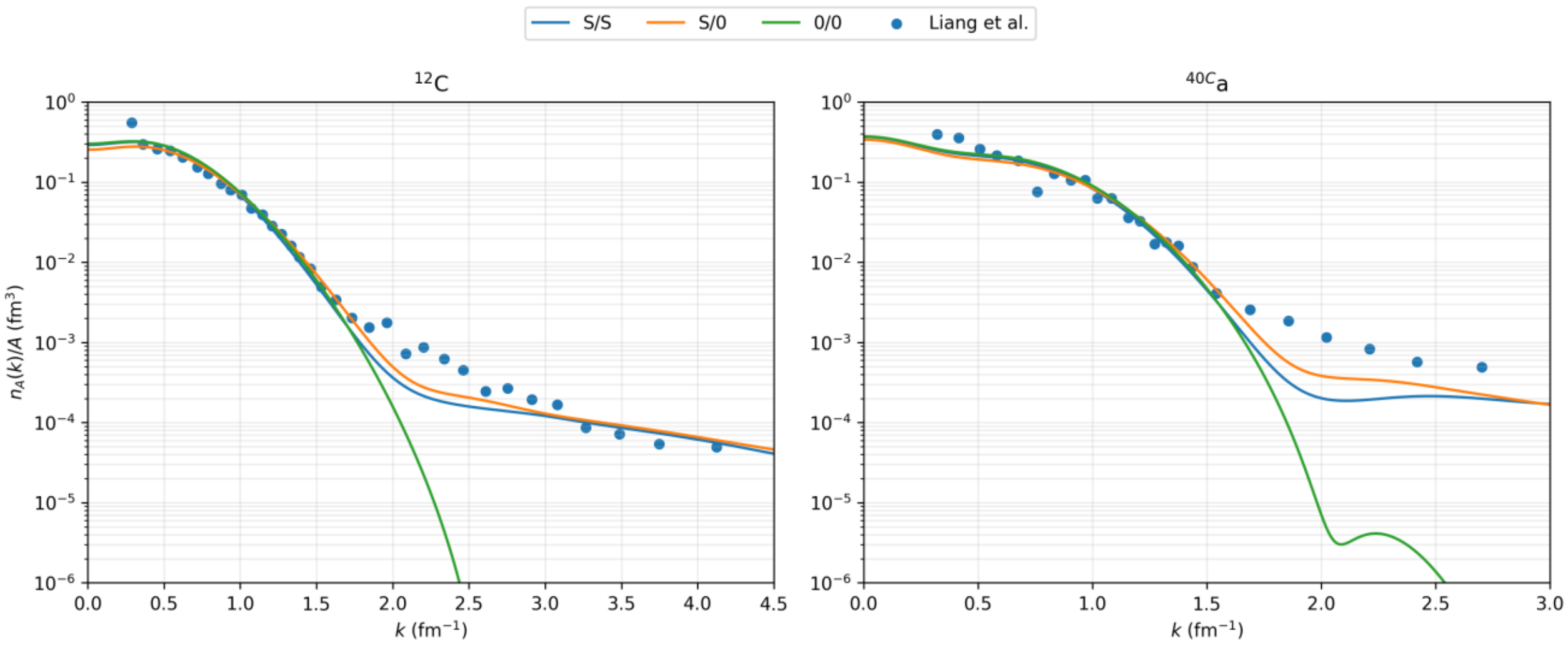

**Figure 8. Corrected absolute per-nucleon momentum distributions for $^{12}$C and $^{40}$Ca. Digitized Liang et al. data [32] are compared with S/S, the diagnostic S/0 calculation, and the consistently fitted no-SRC route 0/0.**

For $^{12}$C, the experimental partial probability over 2.000–4.131 fm$^{-1}$ is 4.629%. The theoretical probabilities over 2.000–5.000 fm$^{-1}$ are 3.694% for S/S, 4.228% for S/0, and 0.0907% for 0/0. For $^{40}$Ca, the experimental partial probability over 2.000–2.709 fm$^{-1}$ is 3.335%, while the corresponding theoretical 2.000–5.000 fm$^{-1}$ values are 4.689%, 5.240%, and 0.0118%. Figure 9 summarizes these integrated strengths. Because the experimental windows end at different momenta, these numbers compare the characteristic several-percent scale and are not used to construct a common-window $\chi^2$.

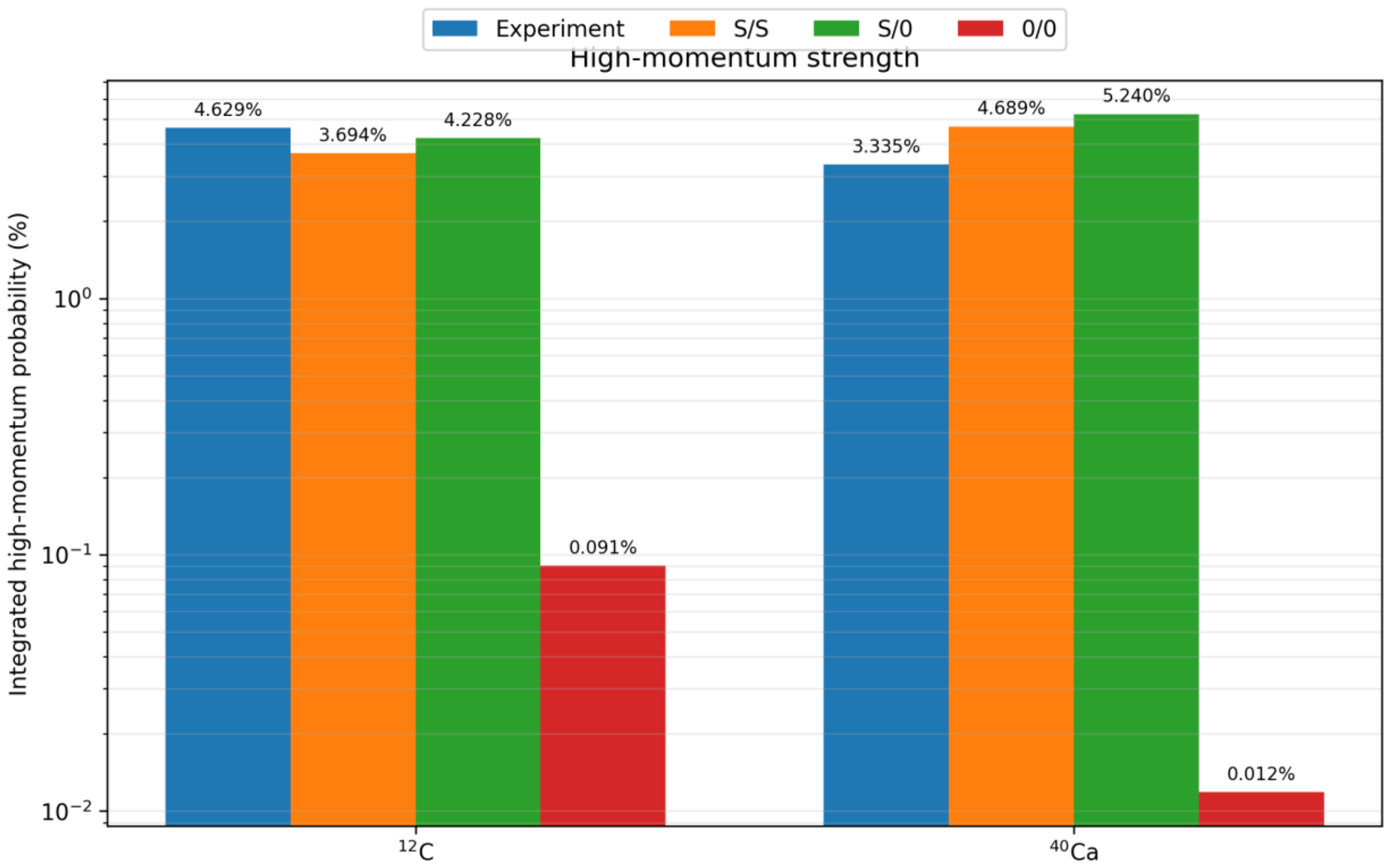

**Figure 9. Integrated high-momentum probabilities from the digitized Liang data and the corrected theoretical calculations. Experimental bars use the available measured windows; theoretical bars use 2≤k≤5 fm⁻¹.**

The direct comparisons lead to a robust distinction between mean-field refitting and explicit correlations. The consistent S/S route and the diagnostic S/0 route both generate several-percent high-momentum strength, while 0/0 does not. Thus refitting is essential for the radius sector, but refitting a smooth no-SRC mean field cannot reproduce the observed tail. S/S remains the physically consistent correlated prediction; S/0 shows that the appearance of the tail itself is tied primarily to the explicit correlation operator rather than to the transferred mean-field parameters.

## 5.8 Numerical and model limitations

The cluster expansion is truncated at first order; the correlation functions are fixed rather than refitted; and the center-of-mass correction is approximate. SRC momentum distributions carry Monte Carlo uncertainty, whereas the no-SRC calculations are effectively deterministic. The genuinely nonlocal OBDM contains orientation-sensitive terms that are invisible to diagonal density and radius checks, so its algebra, hermiticity, particle-number behavior, and independent-seed stability require separate validation. The correlated momentum curves are normalized to the physical proton and neutron numbers after the Fourier transform. The digitized Liang distributions and the theoretical high-k integrals cover different upper momentum limits, so their integrated strengths are treated as scale comparisons and are not combined into a formal $\chi^2$. The annealing plots remain search diagnostics rather than a proof of asymptotic convergence.

# 6. Conclusions and Outlook

Simulated annealing successfully determines Woods–Saxon mean-field parameters within the first-order CBF model with realistic central, spin–isospin, and tensor Short-Range Correlations [5,33]. The correlated and no-SRC calculations must be treated as two parallel, independently optimized routes. Inserting the no-SRC-fitted parameters into the SRC theory increases the radius metric by a factor of 17.49 relative to S/S, demonstrating that model-consistent refitting is essential when explicit SRC are introduced. The 17 fitted nuclei constitute a representative compromise between mass–isospin coverage and correlated-calculation cost, not the full body of available charge-radius data.

After separate optimization, both S/S and 0/0 reproduce the overall charge-radius systematics [25] and give broadly reasonable charge-density and elastic-form-factor shapes relative to the de Vries Fourier–Bessel references [31]. These bulk and coordinate-space observables do not show a universal advantage for the correlated route. Their similarity indicates that a refitted mean field can absorb much of the low-momentum redistribution associated with SRC.

The theoretical re-audit distinguishes the diagonal OBDM from its genuinely nonlocal extension. Retaining both uncorrelated kernels $K\tau$ and $Q\tau$ produces the additional tensor direct contribution $-Q\tau L_i\Gamma$, and the spectator-exchange contraction retains its cubic QQQ orientation term. The direct correction vanishes identically at $r=r'$, so the existing SAM parameters, radii, local densities, and form factors do not require recalculation. With the corrected nonlocal OBDM, the S/S calculation generates the observed several-percent high-momentum scale in $^{12}C$ and $^{40}Ca$, whereas 0/0 gives only a minute tail. The diagnostic S/0 calculation also produces substantial high-k strength, confirming that the tail is generated by explicit SRC rather than by mean-field refitting alone. Future work should provide a fuller common-normalization comparison of proton and neutron momentum distributions, examine higher-order cluster terms, and improve the center-of-mass treatment.

## Acknowledgments

This work was supported in part by the U.S. Department of Energy under Grant No. DE-FG03-93ER40773.

.

# Appendix A. Correlation functions used in the calculation

The correlation functions follow the parameterization in Appendix C of Ref. [6], with parameters from a variational calculation for $^{16}O$ using the Argonne $v_{12}$ interaction [7].

$$g_c(r) = \begin{Bmatrix} 1 + 0.2426r - 5.660r^2 + 125.5r^3 - 791.1r^4 + 2320r^5 \\ -3917r^6 + 4029r^7 - 2494r^8 + 847.2r^9 - 121.8r^{10} \end{Bmatrix} e^{-3.871r^2} \tag{A1}$$

$$f_{t\tau}(r) = \begin{Bmatrix} -0.0143r - 0.1400r^2 - 0.4110r^3 - 0.1863r^4 \\ +2.703r^5 - 3.940r^6 + 2.120r^7 - 0.3311r^8 - 0.0541r^9 \end{Bmatrix} e^{-1.893r^2} \tag{A2}$$

$$f_{s\tau}(r) = \begin{Bmatrix} 0.01037 - 0.01501r + 0.4028r^2 - 6.268r^3 + 38.91r^4 \\ -147.1r^5 + 309.7r^6 - 371.0r^7 + 236.6r^8 - 64.17r^9 \end{Bmatrix} e^{-4.926r^2} \tag{A3}$$

| $g_c(r)$ | $f_{s\tau}(r)$ | $f_{t\tau}(r)$ |
|---|---|---|
| 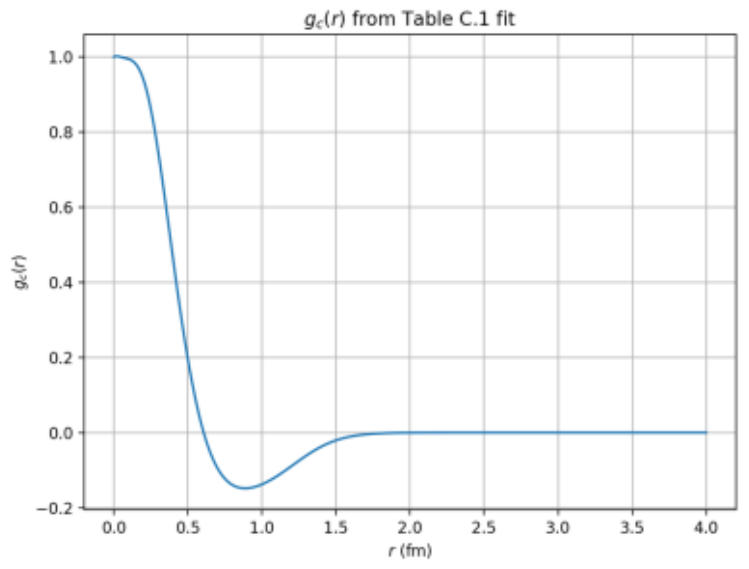 | 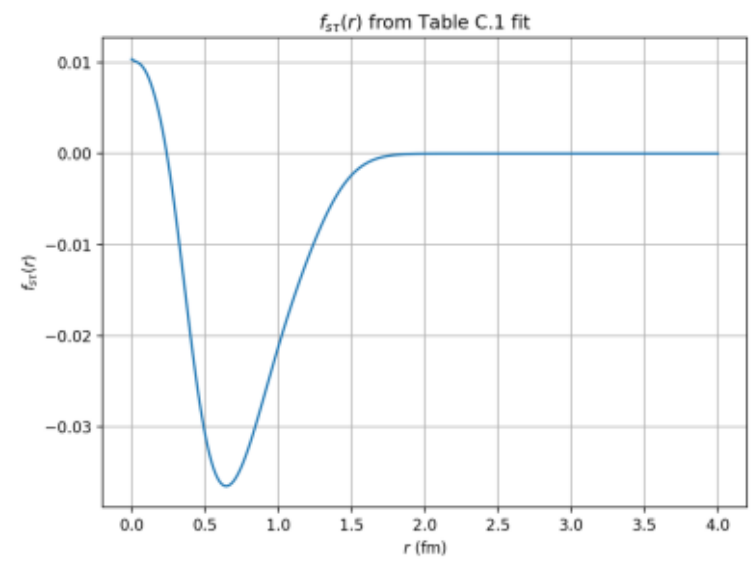 | 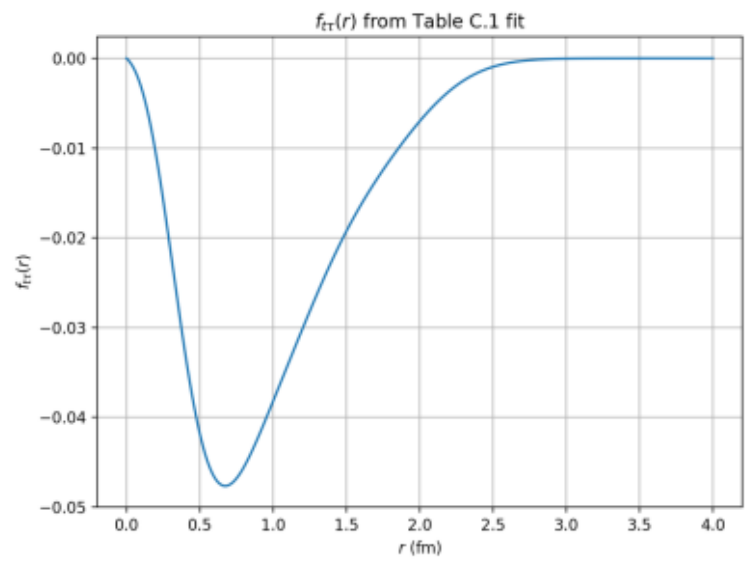 |

*Figure A1. Central, spin–isospin, and tensor correlation functions used in the calculation.*

## Appendix B. Derivation of the separated N≠Z kernels

This appendix gives the derivation that is only summarized in Sec. 2.3. Start from the normalized matrix element of the one-body density operator between the correlated states $F\Phi_0$. Expanding the numerator and denominator to first order in a single correlation link cancels every unlinked term and leaves the four connected Mayer classes shown in Fig. 1: hole/direct, hole/exchange, spectator/direct, and spectator/exchange. Proton and neutron projectors must remain inside the spin–isospin traces until those connected terms have been reduced.

$$\hat{f}(x) = a_x + u_x \sigma_1 \cdot \sigma_2 \tau_1 \cdot \tau_2 + v_x S_{12}(\hat{x}) \tau_1 \cdot \tau_2 \tag{B1}$$

$$\hat{H}(x,y) = \hat{f}(x)\hat{f}(y) - 1 \tag{B2}$$

$$\Xi_{xy} = 3(\hat{x} \cdot \hat{y})^2 - 1 \tag{B3}$$

$$a_x = 1 - g_c(x), u_x = f_{s\tau}(x), v_x = f_{t\tau}(x) \tag{B4}$$

$$P_p = \frac{1+\tau_z}{2}, P_n = \frac{1-\tau_z}{2} \tag{B5}$$

Write the pair operator and its product at the two external coordinates as in Eqs. (B1)–(B5). The central, spin–isospin, and tensor–isospin pieces generate scalar, $\sigma_1 \cdot \sigma_2$, and $S_{12}$ structures. The trace reduction uses Tr $\sigma_i$=Tr $\tau_i$=0, the Pauli-matrix contraction identities, and the tensor product $S_{12}(\hat{x})S_{12}(\hat{x}')$. Its spin-scalar part is collected into $\Xi_{xx'} = 3(\hat{x} \cdot \hat{x}')^2 - 1$. When the active uncorrelated OBDM is genuinely nonlocal, the antisymmetric tensor–tensor part also generates a spin-vector operator proportional to $\sigma \cdot (\hat{x} \times \hat{x}')$. Terms odd in a single spin or isospin operator vanish only after the relevant trace has been completed; they must not be discarded before the species projectors, the two OBDM kernels K and Q, and the exchange ordering are fixed.

For each ordered species pair αβ, the normalized direct and exchange traces are evaluated before any proton–neutron terms are combined:

$$H^D_{\alpha\beta}(x,y) = N^{-1}_{\alpha\beta} Tr_{\sigma,\tau}\left[P_\alpha(1)P_\beta(2)H(x,y)\right), N_{\alpha\beta} = Tr_{\sigma,\tau}\left[P_\alpha(1)P_\beta(2)\right)$$

$$H^E_{\alpha\beta}(x,y) = N^{-1}_{\alpha\beta} Tr_{\sigma,\tau}\left[P_\alpha(1)P_\beta(2)H(x,y)P^\sigma_{12}P^\tau_{12}\right)$$

The exchange trace contains the spin- and isospin-exchange operators

$$P^\sigma_{12} = \frac{1+\sigma_1 \cdot \sigma_2}{2}, P^\tau_{12} = \frac{1+\tau_1 \cdot \tau_2}{2}$$

Together with the Pauli identities

$$(\sigma_1 \cdot \sigma_2)^2 = 3 - 2\sigma_1 \cdot \sigma_2, (\tau_1 \cdot \tau_2)^2 = 3 - 2\tau_1 \cdot \tau_2$$

These traces separate the central–central, spin–isospin, tensor–tensor, and exchange-only mixed central–spin contributions. Direct mixed central–spin terms vanish, but the tensor–tensor trace has both a scalar $H_i$ part and an antisymmetric spin-vector $L_i$ part. The latter contributes only when it is contracted with the Q kernel of a nonlocal active OBDM. Exchange operators additionally make mixed central–spin terms survive with orientation-dependent signs. This is the algebraic origin of the difference between the direct $H_1$–$H_3$/$L_1$–$L_3$ sector and the exchange $H_4$–$H_9$ sector.

For the direct diagrams, the ket and bra retain the same particle ordering. Inserting Pp or Pn on the active line and tracing the internal line gives three inequivalent species traces. Their spin-scalar radial kernels are $H_1$, $H_2$, and $H_3$ in Eqs. (B6)–(B8). $H_1$ and $H_2$ have the same radial polynomial but remain separately labelled because they multiply different proton and neutron internal-line choices; $H_3$ is the mixed-isospin direct trace. The accompanying spin-vector kernels and the full direct matrices are

$$H_1 = a_x a_y - 1 + 3u_x u_y + 3v_x v_y \Xi_{xy} \tag{B6}$$

$$H_2 = a_x a_y - 1 + 3u_x u_y + 3v_x v_y \Xi_{xy} \tag{B7}$$

$$H_3 = 12u_x u_y + 12v_x v_y \Xi_{xy} \tag{B8}$$

$$D(x,x') = \hat{x} \times \hat{x}', \Gamma = C(r,r') \cdot D(x,x') \tag{B8a}$$

$$L_1 = L_2 = 9v(x)v(x')(\hat{x} \cdot \hat{x}'), L_3 = 36v(x)v(x')(\hat{x} \cdot \hat{x}') \tag{B8b}$$

$$M_\mathrm{i} = H_\mathrm{i} I + iL_\mathrm{i}\sigma \cdot D(x,x') \tag{B8c}$$

$$½Tr_s[\rho_0 \tau M_\mathrm{i}] = K\tau H_\mathrm{i} - Q\tau L_\mathrm{i}\Gamma \tag{B8d}$$

The second term in Eq. (B8d) is strictly nonlocal: it vanishes for r′=r because both C and D_xx′ vanish. It therefore changes the momentum distribution without altering the fitted radii or the local density and form-factor calculations.

For exchange diagrams, the bra and ket spin–isospin labels are interchanged before the trace. The exchange operator changes both the Pauli traces and the OBDM ordering. Same-species exchange produces $H_4$ and $H_5$. The two distinct mixed-species orientations produce $H_6$,$H_7$ and $H_8$,$H_9$. They are not interchangeable for N≠Z because $\rho_0$p(r,y)$\rho_0$n(y,r′) and $\rho_0$n(r,y)$\rho_0$p(y,r′) are different nonlocal products. Evaluating these six ordered traces gives Eqs. (B9)–(B14).

$$H_4 = a_x a_y - 1 - a_x u_y - u_x a_y + u_x u_y + 2v_x v_y \Xi_{xy} \tag{B9}$$

$$H_5 = a_x u_y + u_x a_y - 2u_x u_y + v_x v_y \Xi_{xy} \tag{B10}$$

$$H_6 = -a_x u_y - 5u_x u_y - 2v_x v_y \Xi_{xy} \tag{B11}$$

$$H_7 = a_x u_y + 2u_x u_y - v_x v_y \Xi_{xy} \tag{B12}$$

$$H_8 = -u_x a_y - 5u_x u_y - 2v_x v_y \Xi_{xy} \tag{B13}$$

$$H_9 = u_x a_y + 2u_x u_y - v_x v_y \Xi_{xy} \tag{B14}$$

The labels $H_6/H_8$ and $H_7/H_9$ are retained separately precisely to preserve the pn and np orientations. For local density and radius observables, the external coordinates are set equal only after this ordered reduction. Then x=y, $\Xi_{xx}$=2, and the nonlocal kernels reduce to the one-vector kernels $h_1,\ldots,h_9$ in Eqs. (B15)–(B24).

$$h_i(d) = H_i(\mathbf{x}, \mathbf{x}), \qquad d = |\mathbf{x}|, \qquad \Xi_{xx} = 2 \tag{B15}$$

$$h_1 = a^2 - 1 + 3u^2 + 6v^2 \tag{B16}$$

$$h_2 = a^2 - 1 + 3u^2 + 6v^2 \tag{B17}$$

$$h_3 = 12u^2 + 24v^2 \tag{B18}$$

$$h_4 = a^2 - 1 - 2au + u^2 + 4v^2 \tag{B19}$$

$$h_5 = 2au - 2u^2 + 2v^2 \tag{B20}$$

$$h_6 = -au - 5u^2 - 4v^2 \tag{B21}$$

$$h_7 = au + 2u^2 - 2v^2 \tag{B22}$$

$$h_8 = -au - 5u^2 - 4v^2 \tag{B23}$$

$$h_9 = au + 2u^2 - 2v^2 \tag{B24}$$

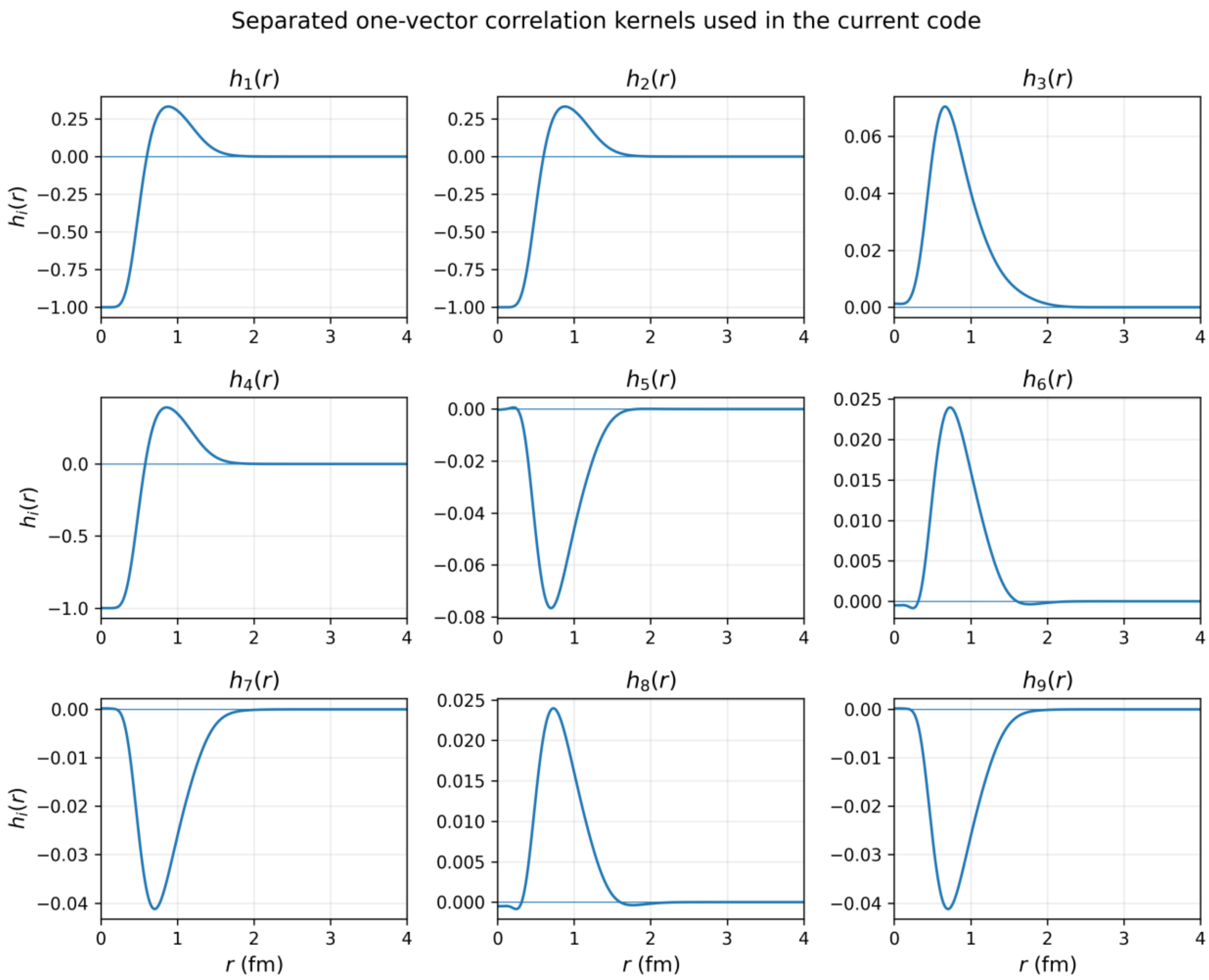

Figure B1. Separated one-vector kernels $h_1,\ldots,h_9$ used in the diagonal OBDM, density, and radius calculations. The combined figure is retained in Appendix B rather than the main article.

The role of each label in the OBDM reduction is summarized below.

| Kernel(s) | Trace class | OBDM role |
|---|---|---|
| $H_1$,$H_2$ | direct, species preserving | same radial trace retained separately for the proton and neutron internal-line choices |
| $H_3$ | direct, mixed isospin | different-isospin direct contribution |
| $H_4$ | exchange, same species | orientation-sensitive two-line contraction $X_{pp}$ (or $X_{nn}$) |
| $H_5$ | exchange, same species scalar | product of spin-scalar K amplitudes |
| $H_6$,$H_7$ | exchange, pn orientation | active proton followed by neutron internal ordering |
| $H_8$,$H_9$ | exchange, np orientation | reversed neutron–proton ordering; distinct for $N \neq Z$ |
| $L_1$,$L_2$,$L_3$ | direct, nonlocal spin-vector | tensor–tensor contribution coupled to $Q\tau$; vanishes identically for $r'=r$ |

The OBDM contractions in Eqs. (B25)–(B27) keep the spin sums explicit. X is a two-line contraction and Y is a three-line contraction. To make the implementation requirement explicit, write each spin matrix as $M_j=K_jI+iQ_j\sigma\cdot u_j$. The spin-scalar part of the three-line product is

$$X_{\alpha\beta;ss'}(a,b,c) = \sum_{s_1} \rho_{0\alpha;ss_1}(a,b)\rho_{0\beta;s_1s'}(b,c) \tag{B25}$$

$$\tilde{X}_{\alpha\gamma;ss'}(a,b;c,d) = \sum_{s_1} \rho_{0\alpha;ss_1}(a,b)\rho_{0\gamma;s_1s'}(c,d) \tag{B26}$$

$$Y_{\alpha\beta\gamma;ss'}(a,b,c,d) = \sum_{s_1s_2} \rho_{0\alpha;ss_1}(a,b)\rho_{0\beta;s_1s_2}(b,c) \times \rho_{0\gamma;s_2s'}(c,d) \tag{B27}$$

$$½Tr(M_1M_2M_3) = K_1K_2K_3 - \Sigma cycK_1Q_2Q_3(u_2 \cdot u_3) + Q_1Q_2Q_3u_1 \cdot (u_2 \times u_3) \tag{B27a}$$

Here Σcyc denotes the three cyclic KQQ terms. The final cubic QQQ term is required in the spectator/exchange class. In the coordinate implementation, its triple product is evaluated through the corresponding orientation determinants. Substituting the complete X and Y contractions into the four connected diagram classes gives the proton OBDM in Eqs. (B28)–(B32); the neutron expression follows by the full p↔n interchange in Eq. (B33), including reversal of pn and np orientations.

The proton OBDM is organized as the uncorrelated term plus hole/direct $A_{\mathrm{p}}$, hole/exchange $B_{\mathrm{p}}$, spectator/direct $C_{\mathrm{p}}$, and spectator/exchange $D_{\mathrm{p}}$:

$$\rho_p = \rho_{0p} + A_p - B_p - C_p + D_p \tag{B28}$$

$$A_{\mathrm{p}} = 2\int d^3y K_{\mathrm{p}}(y,y)[\rho_{0\mathrm{p}}M_1] + 2\int d^3y K_{\mathrm{n}}(y,y)[\rho_{0\mathrm{p}}M_2] + 2\int d^3y K_{\mathrm{p}}(y,y)[\rho_{0\mathrm{n}}M_3] \tag{B29}$$

$$B_p = \int d^3y\big(X_{pp}H_4 + 8K_pK_pH_5\big) + \int d^3y\big(2X_{pn}H_6 + 16K_pK_nH_7\big) + \int d^3y\big(2X_{np}H_8 + 16K_nK_pH_9\big) \tag{B30}$$

$$C_p = 2\int d^3y d^3z X_{pp}\{K_ph_1 + K_n(h_2 + h_3)\} \tag{B31}$$

For the spectator/exchange class, the correlation link joins the two integrated spectator coordinates. With the abbreviation dΓ for the coordinate integrations and the indicated OBDM products, the result is

$$D_p = \int d\Gamma Y_{ppp}h_4 + 4\int d\Gamma \tilde{X}_{pp}K_ph_5 + 2\int d\Gamma Y_{pnp}(h_6 + h_8) + 8\int d\Gamma \tilde{X}_{pp}K_n(h_7 + h_9) \tag{B32}$$

$$A_n, B_n, C_n, D_n = \big(A_p, B_p, C_p, D_p\big)\big|_{p\leftrightarrow n} \tag{B33}$$

The N=Z check is performed only after the separated expressions have been assembled. Setting $K_p=K_n$, $Q_p=Q_n$, and $\rho_p=\rho_n$ makes the two mixed-species exchange orientations equal. The scalar $H_i$ sector then recombines into the conventional spin- and isospin-summed expressions, while the nonlocal direct $L_i$ term must still be retained whenever

Q≠0 and r′≠r. Adding the proton and neutron OBDMs reproduces the diagonal benchmark of Appendix C and the full nonlocal result when the Q-dependent direct term is included.

$$K_p = K_n, Q_p = Q_n, \rho_p = \rho_n \quad \Rightarrow N = Z \text{spin- and isospin-summed limit} \tag{B34}$$

The numerical coefficients in Eqs. (B30) and (B32) belong to different diagram topologies. In Bp, the scalar K products represent two spin-summed one-body lines and therefore carry the factors 8 and 16 shown in Eq. (B30). In Dp, the three-line spectator/exchange contraction and its symmetry factor give the factors 4 and 8 in Eq. (B32). The complete Y contraction includes KKK, all three KQQ terms, and the QQQ orientation term of Eq. (B27a). These are not alternative Monte Carlo estimators. Importance sampling divides each complete physical integrand by its proposal density and leaves every kernel and combinatorial coefficient unchanged. The old compressed B/D/E/T combinations are used only for N=Z checks or proposal-envelope design, never as replacements for the separated N≠Z formulas.

## Appendix C. Spin- and isospin-summed N=Z reduction

This appendix is a benchmark reduction of the spin-scalar and diagonal limits. Its formulas are fully spin- and isospin-summed, as in Appendix C of Ref. [5], and they are not the working formulas for a general N≠Z nucleus. In the equal-species limit $K_p$=$K_n$, $Q_p$=$Q_n$, and $\rho_p$=$\rho_n$, the pn and np exchange contractions become identical. For local observables, r′=r also makes the direct Q-dependent contribution vanish exactly. For a finite-separation OBDM with Q≠0, however, the scalar benchmark must be supplemented by Eqs. (23)–(27) or (B8a)–(B8d).

Using the auxiliary N=Z notation f for the central channel and g for the common spin–isospin/tensor channel of the simplified $f_3$ reduction, the spin- and isospin-summed scalar direct and exchange kernels are

$$H_D(\mathbf{x},\mathbf{y}) = 16\big[\big(f_x f_y - 1\big) + 27 g_x g_y (\hat{\mathbf{x}}\cdot\hat{\mathbf{y}})^2\big]. \tag{C1}$$

$$H_E(\mathbf{x},\mathbf{y}) = 4\big\{\big(f_x f_y - 1\big) + 27 g_x g_y [1 - 2(\hat{\mathbf{x}}\cdot\hat{\mathbf{y}})^2] + 9\big(f_x g_y + f_y g_x\big).\big\} \tag{C2}$$

where the angular factor is retained explicitly. The corresponding diagonal one-vector limits are

$$h_D(r) = 16[(f^2(r) - 1) + 27 g^2(r)]. \tag{C3}$$

$$h_E(r) = 4[(f^2(r) - 1) - 27 g^2(r) + 18 f(r) g(r)]. \tag{C4}$$

The scalar reduction from Appendix B proceeds at the contraction level: $H_1$ and $H_2$ merge after the proton and neutron internal-line choices become identical; $H_3$ supplies the mixed-isospin direct part; $H_6$ and $H_8$ become equal, as do $H_7$ and $H_9$; and $H_4$/$H_5$ provide the same-species exchange pieces. After adding proton and neutron OBDMs and restoring their spin multiplicities, these grouped scalar terms reproduce Eqs. (C1)–(C4). The same grouping of $h_1$–$h_9$ gives the local formulas. This equality checks the diagonal implementation but does not test the off-diagonal $Q\tau L_i\Gamma$ contribution, which is retained separately in the momentum calculation.

## Appendix D. Monte Carlo importance sampling

The first-order cluster expansion contains high-dimensional integrals, so repeated deterministic quadrature is not practical inside the SRC simulated-annealing fit. The numerical method is ordinary importance sampling [29,30]. For a collective coordinate x and a normalized proposal p(x), the physical integrand is divided by p(x) and averaged as written in Eqs. (D1)–(D4). The proposal changes the variance, not the physical kernel coefficients.

$$I = \int f(x)dx \tag{D1}$$

$$I = \int p(x)\left[\frac{f(x)}{p(x)}\right]dx \tag{D2}$$

$$I_{MC} = \frac{1}{N}\sum_{i=1}^{N}\frac{f(x_i)}{p(x_i)} \tag{D3}$$

$$\sigma_I^2 = \frac{1}{N}\left\{\int p(x)\left[\frac{f(x)}{p(x)}\right]^2 dx - I^2\right\} \tag{D4}$$

The short-range proposal centered on an active point is

$$p_\beta(\mathbf{r}_1|\mathbf{r}_0) \propto \rho_\tau^2(r_1)\exp[-\beta|\mathbf{r}_1 - \mathbf{r}_0|^2] \tag{D5}$$

$$V_X(\beta) \propto \beta^{-3/2}\int d^3r G_X^2(r)e^{\beta r^2} \tag{D6}$$

$$\frac{dlnV_X}{d\beta} = 0 \tag{D7}$$

$$-\frac{3}{2\beta} + \frac{\int d^3r r^2 G_X^2 e^{\beta r^2}}{\int d^3r G_X^2 e^{\beta r^2}} = 0 \tag{D8}$$

$$\beta_X = \frac{3}{2\langle r^2\rangle_X(\beta_X)} \tag{D9}$$

$$\langle r^2\rangle_X(\beta) = \frac{\int d^3r r^2 G_X^2 e^{\beta r^2}}{\int d^3r G_X^2 e^{\beta r^2}} \tag{D10}$$

$$\beta_X^{(n+1)} = \frac{3}{2\langle r^2\rangle_X\left(\beta_X^{(n)}\right)} \tag{D11}$$

$$b_B = 1.28455 fm^{-2} \tag{D12}$$

$$b_D = 1.24320 fm^{-2} \tag{D13}$$

$$b_E = 1.91469 fm^{-2} \tag{D14}$$

$$b_b = b_B, b_d = b_D, b_e = b_E \tag{D15}$$

The labels in Eqs. (D12)–(D15) denote proposal-width groups only. $b_B$ is used for the B-type direct envelope, $b_D$ for the D-type direct envelope, and $b_E$ as a common exchange envelope. These auxiliary Gaussian widths select efficient samples; every accepted sample is evaluated with the exact separated $H_i$ or $h_i$ integrand.

$$p_\tau = \int 4\pi r^2 \rho_\tau(r) dr \tag{D16}$$

$$\langle r^2 \rangle_\tau = \frac{\int 4\pi r^4 \rho_\tau(r) dr}{\int 4\pi r^2 \rho_\tau(r) dr} \tag{D17}$$

$$L_\tau = \sum_{nlj\in\tau}^{occ} \frac{l(l+1)\left(l+s_{lj}\right)}{2} \tag{D18}$$

$$s_{lj} = 1(j = l + 1/2), s_{lj} = 0(j = l - 1/2) \tag{D19}$$

$$b_{L,\tau} = \frac{0.10 L_\tau}{p_\tau \langle r^2 \rangle_\tau} \tag{D20}$$

The radial cumulative table used for inverse-transform sampling is

$$N_b(r_0, r) = \int_0^r d\, s \frac{\pi \rho_\tau^2(s) s}{b r_0} \{e^{-b(s-r_0)^2} - e^{-b(s+r_0)^2}\}, \tag{D21}$$

$$p_b(r_0) = N_b(r_0, \infty) \tag{D22}$$

$$P(c) \propto e^{2br_0 r_1 c}, c = cos\theta \tag{D23}$$

$$p_0 = 2br_0 r_1 \tag{D24}$$

$$c = \frac{\ln[x e^{p_0} + (1-x) e^{-p_0}]}{p_0} \tag{D25}$$

$$c = 2x - 1 \qquad (|p_0| \text{ small}) \tag{D26}$$

$$P_r(r) \propto \rho_p(r) r^4 \tag{D27}$$

The $r^4$ factor contains both the radial volume element and the $r^2$ observable weight. If the spectator anchor is sampled from $\rho_\tau^2$, the returned normalization is corrected as

$$p_{10} \to p_{10}/\rho_\tau\,(r_{10}) \tag{D28}$$

$$\hbar_1(d) = h_1(d) e^{b_B d^2} \tag{D29}$$

$$\tilde{h}_{23}(d) = [h_2(d) + h_3(d)] e^{b_D d^2} \tag{D30}$$

The two orientation-sensitive exchange products retained for $N \neq Z$ are

$$R_{pn}^{(2)} \propto \sum_{ss_1} \rho_{0p;ss_1}\,(r, r_1) \rho_{0n;s_1 s}(r_1, r') \tag{D31}$$

$$R_{np}^{(2)} \propto \sum_{ss_1} \rho_{0n;ss_1}\,(r, r_1) \rho_{0p;s_1 s}(r_1, r') \tag{D32}$$

$$G_{\mathrm{ex}}^{(w)}(r) \sim |h_4| + 4|h_5| + |2h_6 + 2h_8| + 8|h_7 + h_9| \tag{D33}$$

Equation (D33) defines only an envelope for selecting an efficient common exchange width. The estimator evaluates the exact pn and np contractions and divides the complete integrand by the normalized proposal density. A factor such as 1/ρτ is mathematically legitimate when the corresponding sample is drawn from a proposal with zero probability wherever ρτ=0; the explicit zero-density guard is retained only for finite-grid underflow and inverse-CDF edge cases. Species-matched proposal tables are used where required, and the same random block is reused across the momentum half-separation grid. The final density and form-factor curves are averages of three independent 300,000-sample unsmoothed runs. Sample-size and independent-seed calculations are numerical stability checks; no claim of asymptotic convergence is made.